\documentstyle[11pt,paspconf,epsfig]{article}

\begin{document}

\title{Gas-rich LSB Galaxies -- Progenitors of Blue Compact Dwarfs?}

\author{John J. Salzer}
\affil{Astronomy Department, Wesleyan University, Middletown, CT 06459}

\author{Stuart A. Norton}
\affil{Lick Observatory, Univ. California Santa Cruz, Santa Cruz, CA 95064}


\begin{abstract}
We analyze deep CCD images of nearby Blue Compact Dwarf (BCD) galaxies
in an attempt to understand the nature of the progenitors which
are hosting the current burst of star formation.  In particular, we ask
whether BCDs are hosted by normal or low-surface-brightness dI galaxies.
We conclude that BCDs are in fact hosted by gas-rich galaxies which 
populate the extreme high-central-mass-density end of the dwarf galaxy 
distribution.  Such galaxies are predisposed to having numerous strong
bursts of star formation in their central regions.  In this picture,
BCDs can only occur in the minority of dwarf galaxies, rather than
being a common phase experienced by all gas-rich dwarfs.
\end{abstract}


\keywords{Galaxies: Evolution; Galaxies: Irregular; Galaxies: Starburst}

\section{Introduction}

If one wished to adopt the simplest scheme for classifying dwarf 
galaxies, most could be lumped into one of two categories. The first 
would be the dwarf irregulars (dIs), which can be characterized
as having plenty of gas and usually some level of recent or
current star formation.  The other would be dwarf ellipticals
(dEs), which by comparison to dIs have little if any gas, and
usually no significant recent star formation (although exceptions
to this latter point certainly exist).  Interestingly, both types
of dwarfs tend to have surface brightness distributions that are
well fit by simple exponential profiles.  Structural parameters
derived from surface photometry of both dIs and dEs show a
large range of values: galaxies of both types are observed 
with both high and low central surface brightnesses ($\mu_0$),
and both large and small exponential scale-lengths ($\alpha$).
In fact, the two types overlap completely in the the $\mu_0$ --
$\alpha$ plane.

A group of low-luminosity galaxies which do not readily lend
themselves to classification in the above scheme are the blue
compact dwarfs (BCDs).  These are dwarf galaxies which are currently
undergoing an extremely strong burst of star formation, such that
the optical appearance of the galaxy is dominated by the energy
output of the young stars.  In some cases the starburst is so
dominant that the presence of an underlying older population of
stars is not clearly evident.

Given their extreme nature, it has been difficult to determine
with any confidence the type of galaxy that typically hosts BCDs. 
Since BCDs are usually observed to be gas rich (Thaun \& Martin 
1983, Salzer et al. 1999a), the most common assumption is that BCDs 
represent bursts of star formation occurring in dI galaxies.  But 
can any dI initiate a large star-formation episode and appear as 
a BCD?  This is a key question, since it impacts our picture of how 
dwarf galaxies evolve.  Is the BCD phenomenon a stage of galaxy 
evolution common to all gas-rich dwarfs?  We attempt to answer 
this question in the current study.

\section{Relevant Facts about BCDs}

Before attempting to address the question posed in the previous
section, we review some of the relevant characteristics of BCDs:

$\bullet$ {\bf Optical appearance dominated by light from starburst}.
As mentioned above, BCDs are dwarf galaxies whose optical light
output is dominated by the energy released by the starburst 
component.  This includes the light from the young O and B stars, plus
the nebular emission (both line and continuum) which represents
reprocessed UV radiation from the same massive stars.  The latter can 
be a major contributor to the broad-band fluxes measured in BCDs, in
extreme cases exceeding the light of the stellar component in the
optical.  This characteristic of BCDs has made classification of
the underlying host galaxy all but impossible in many cases.
\medskip

$\bullet$ {\bf Very intense nebular spectra}.  Spectra of BCDs are
dominated by nebular emission lines.  In most cases, strong 
recombination lines of H and He obliterate any stellar absorption 
lines which may be present.  In addition, the nebular continuum
combined with the relatively featureless continua of the O and B
stars acts to effectively hide the presence of lines from the older
stars of the host galaxy.  Consequently, spectroscopy yields very 
little information regarding the stellar content of the BCD host 
galaxy.  On the other hand, the nebular spectra {\it do} allow for 
the accurate determination of the abundances in the ionized gas.
\medskip

$\bullet$ {\bf Gas rich}.  The typical BCD contains a large amount
of HI gas.  The mean value of M$_{HI}$/M$_{tot}$ for a sample
of 122 BCDs is 0.16 (Salzer et al. 1999a).  Figure 1 shows the 
distribution of M$_{HI}$/L$_B$ for this sample, plotted vs.
absolute magnitude.  The open symbols represent average values for 
spiral and irregular galaxies taken from the literature.  
A statistical correction to the luminosities and 
mass-to-light ratios of the BCDs has been applied to account for
the fact that their luminosities are elevated by an average of
0.75 B magnitudes due to the starburst (see below).  Thus, the figure 
shows where the {\bf host galaxies of the BCDs} would lie in this diagram
if the starburst component were removed.  On average, the BCD hosts
have a factor of $\sim$2 higher HI gas mass at a given absolute magnitude
than do the more quiescent irregular galaxies with which they are
compared.  To be sure, there are some BCDs with low HI gas content,
but on the whole, BCDs are quite gas rich.  

\begin{figure}[t]
\begin{center}
\leavevmode
\epsfverbosetrue
\epsfxsize=0.99\textwidth
\epsfbox{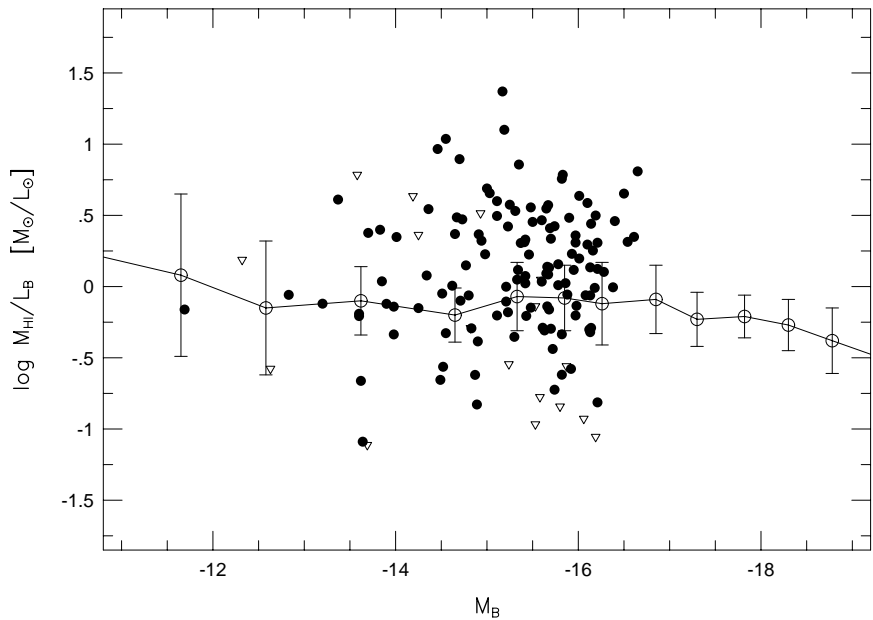}
\end{center}
\vskip -0.4in
\caption[]{The HI content of BCDs.  Here we plot the HI mass to blue-light 
ratio vs. absolute magnitude for a large sample of BCDs taken from
Salzer et al. 1999a.  The BCDs are shown as filled circles, while open
triangles show upper limits for the BCDs not detected in HI.  The connected
open circles are the mean values of M$_{HI}$/L$_B$ for a sample of spiral
and irregular galaxies taken from the literature.  After correcting the
BCDs for the excess light due to their starburst, they are seen to mostly
lie at or above the location of the normal galaxies in this diagram.}
\label{fig:plot1}
\end{figure}

The presence of many BCD hosts lying above the trend set by the 
comparison sample could be interpreted in two ways: either they have 
unusually high HI masses, or they have normal HI masses but unusually 
low luminosities.  The presence of a population of galaxies in the
upper portion of this diagram, with no counterpart in the comparison
sample, suggests the possibility that these galaxies would be
difficult to detect in typical galaxy surveys during their quiescent
phases.  This led us to suspect originally that the BCD hosts might
in fact be LSB dwarfs.
\medskip

$\bullet$ {\bf Most not bursting for the first time}.
The dominance of the starburst led a number of authors to suggest 
that the observed burst of star formation in some BCDs represents 
the first episode of star formation in these objects (e.g., Searle \& 
Sargent 1972).  However, more recent imaging studies have shown fairly 
convincingly that nearly all BCDs do possess an older, underlying 
population of stars (e.g., Papaderos et al. 1996a,b; Telles \& 
Terlevich 1997).  There are a few galaxies, such as I Zw 18 and SBS 
0335-04, for which a strong case can be made that the current starburst 
represents the first major episode of star formation (e.g., Thuan et al. 
1997), but these are the exception rather than the rule.  Although
it is virtually impossible to assign ages to the underlying population,
one can say with some confidence that most BCDs began forming stars
long ago.
\medskip

$\bullet$ {\bf Burst strengths not extreme}.  Because of the optical
dominance of the starburst component in BCDs, it is often assumed that
the current burst involves a large fraction of the mass of the galaxy,
and that it has elevated the brightness of the galaxy by a large
amount (estimated at 3--5 magnitudes by some authors).  However, this
turns out to be an overestimate.  Modeling of the starbursts in over a 
dozen extreme BCDs shows that, on average, only a few percent of the 
available HI gas is being used in the current starburst (Salzer et al. 
1999b).  Further, this work shows that the average B-band luminosity 
enhancement due to the starburst (both stellar and nebular contributions)
is only 0.75 magnitudes.  Thus, the ``bursts" in BCDs are not such
extreme events as one might think.
\medskip

\section{Some Recent Key Results}

Three recent results have played a significant role in reshaping
our view of the nature of BCD host galaxies (or at least in how
we might interpret the available data).

The first of these are theoretical studies which attempt to 
account for the fate of the ISM in dwarf galaxies which experience
a major starburst.  Early work suggested that even modest numbers
of supernovae were enough to completely remove the ambient gas
in small galaxies.  One implication of this was that BCDs
would lose all of their gas following the starburst event, and
after 1-2 Gyr resemble dwarf ellipticals.  However, more
recent studies (DeYoung \& Heckman 1994; MacLow \& Ferrarra 1998; 
Brighenti \& D'Ercole 1999) have
come to the opposite conclusion: only for extremely low-mass
galaxies does a starburst remove all of the gas.  These new
simulations suggest that the hot SN ejecta (including most of the 
metals produced in the high mass stars) will escape, while the bulk
of the colder ISM will remain.  If correct, these new studies
change drastically our view of post-BCDs.  This result is also
consistent with the picture that BCDs have had previous star formation.
If dwarf galaxies lost their gas easily due to SN outflows, then
they would have trouble creating additional generations of stars.
Since most BCDs are known to possess at least two generations of
stars (and perhaps many more), the ability to retain their gas is
obviously crucial.

Another recent finding that may play a major role in our understanding
of the BCD phenomenon is that BCDs not only have more HI gas than
comparable-sized dIs, but the gas is {\it more centrally concentrated}.
In a recent paper by van Zee et al. (1998), the azimuthally-averaged 
HI distributions for 8 BCDs were compared to those for a similar
number of dIs.  The BCDs tend to have strongly peaked gas distributions, 
i.e., a large reservoir of HI in the central portions of the galaxies.  
We believe that this is related to the presence of the starbursts in 
the BCDs, and in fact may be a necessary condition for the occurrence 
of a strong, sustained star-formation episode like that seen in BCDs.

The third result which has had a major impact on our view of BCD
hosts, and which as the primary motivation for carrying out the
current study, is the recent work by Papaderos et al. (1996a,b) and
Telles \& Terlevich (1997).  These studies investigate the surface
brightness distributions of BCDs, and shed new light on the nature
of the host galaxies of BCDs.  In particular, Papaderos et al. 
utilized surface photometry of BCDs out to faint surface brightness 
levels which allowed them to study both the distribution of the 
starburst light as well as the light from the underlying host galaxy.
Among their results was the suggestion that the host galaxies of
BCDs have significantly different characteristics than other, more
normal, dwarf galaxies.  Their success gave us the incentive to 
carry out a similar analysis on a large number of existing BCD images 
obtained previously for other purposes (Salzer \& Elston 1992; 
Salzer et al. 1999b). 

\section{Structural Parameters of BCD Host Galaxies}

We have carried out detailed surface photometry using deep B-band 
CCD images for a sample of 18 BCDs and 11 dIs.  The BCDs were 
selected from a variety of survey lists, and were chosen to represent 
the subsample of dwarf star-forming galaxies with the most extreme 
properties (i.e., the most intense star-formation events).  The dI 
galaxies were analyzed as a comparison sample.  These galaxies are 
part of a separate study of the properties of nearby dwarf galaxies 
known to exhibit numerous holes in their HI distributions (see Rhode 
et al. 1999, this volume).  Additional comparison dIs were taken 
from the study by Patterson \& Thuan (1996).  Both the BCD and dI 
samples were limited to galaxies with M$_B$ $>$ $-$17.

Since the BCDs are dominated by a (usually) central starburst, 
isophotal fitting is a tricky business.  In general, it is not trivial 
to distinguish between light from the underlying host galaxy and light 
from the starburst.  However, since we were not particularly interested 
in the starburst region, we adopted the following simple approach.  
First, we allowed our isophote-fitting software to fit the entire galaxy
with no preset restrictions (i.e., we did not exclude the starburst
region).  We then used the radial brightness
profiles derived from the isophote fitting to determine the structural
parameters of the galaxy: central surface brightness ($\mu_o$) and
exponential scale length ($\alpha$).  We assumed that the underlying
host galaxy could be represented by a single exponential profile.  The 
exponential disk profile was fit {\it only in the outer portions of the
galaxy, well outside the radius occupied by the starburst}.  We used 
H$\alpha$ images of each galaxy to define where the star formation was 
occurring.  In this way, we can be fairly confident that the derived
parameters represent those of the underlying galaxy, and are not affected
by the intense starburst.  For consistency, we carried out fits to the
radial brightness profiles of the comparison dIs in the exact same way.
Only the outer portions of the dI profiles were used to determine the
structural fits.  Complete details of our analysis are given in Norton
\& Salzer (1999).

\begin{figure}[t]
\begin{center}
\leavevmode
\epsfverbosetrue
\epsfxsize=0.99\textwidth
\epsfbox{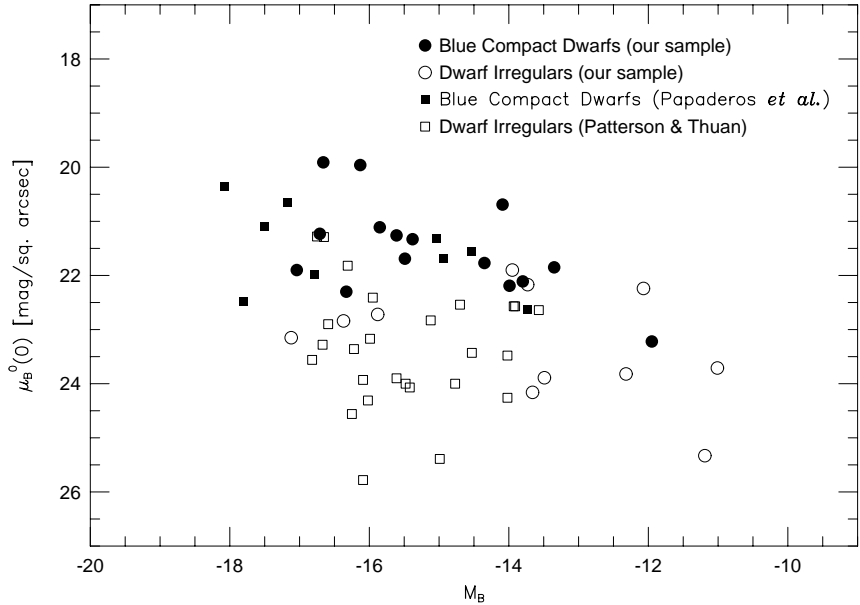}
\end{center}
\vskip -0.4in
\caption[]{Plot of central surface brightness vs. absolute
magnitude for our sample of BCDs and dIs.  Also plotted are BCDs from
Papaderos et al. and dIs from Patterson \& Thuan.  The BCDs have 
systematically higher central surface brightnesses than do the dIs.}
\label{fig:plot2}
\end{figure}

\begin{figure}[t]
\begin{center}
\leavevmode
\epsfverbosetrue
\epsfxsize=0.99\textwidth
\epsfbox{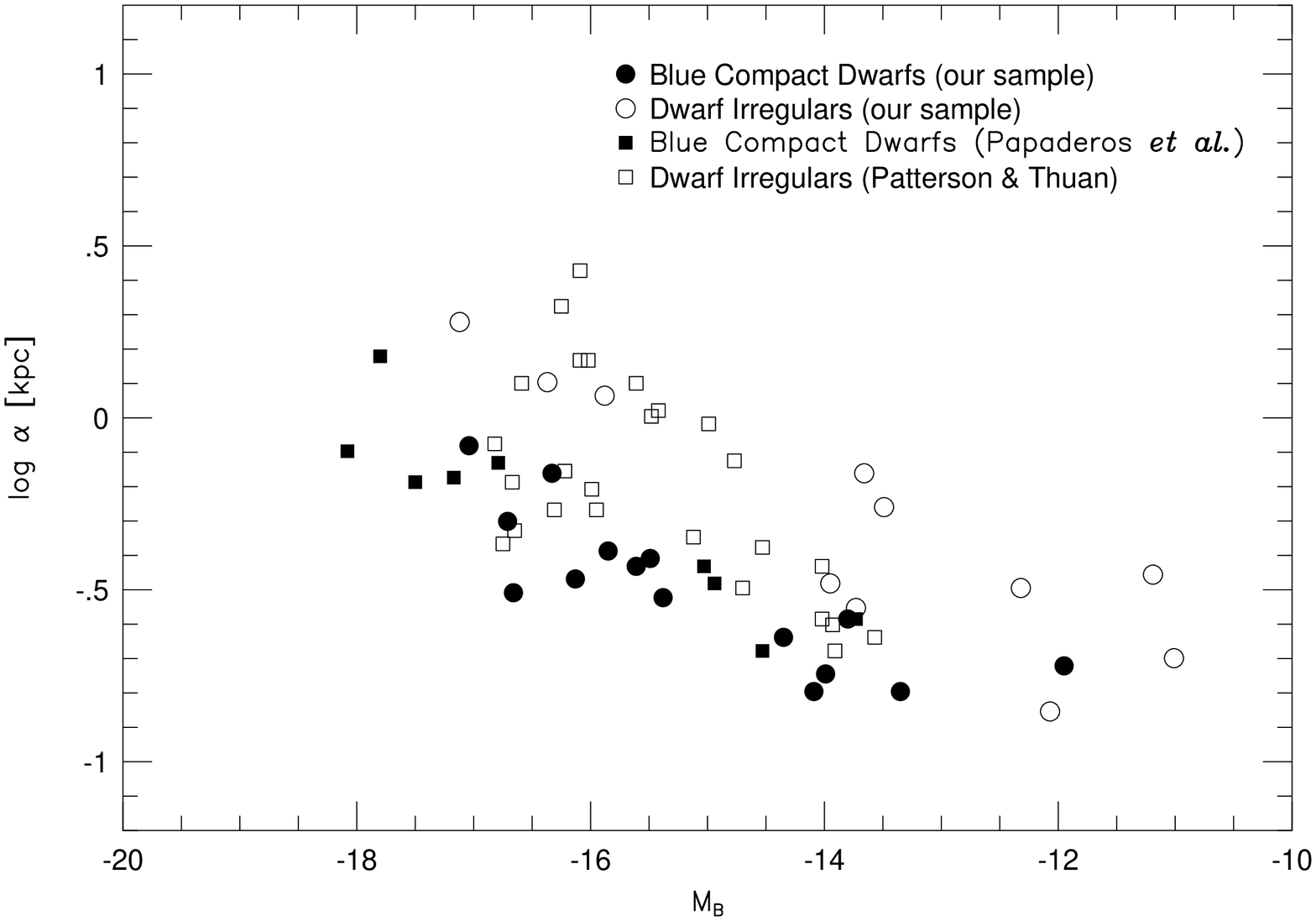}
\end{center}
\vskip -0.4in
\caption[]{Plot of the exponential disk scale length ($\alpha$) vs.
absolute magnitude for our sample of BCDs and dIs.  Also plotted are 
BCDs from Papaderos et al. and dIs from Patterson \& Thuan.  The BCDs
tend to possess smaller disk scale lengths than the dIs, indicating a
more compact distribution for the underlying stars.}
\label{fig:plot3}
\end{figure}

The results of our structural parameter determinations are shown in
Figures 2 and 3.  Figure 2 plots the extrapolated central surface 
brightness vs. absolute
magnitude, while Figure 3 shows disk scale length plotted against absolute
magnitude.  Also plotted are similar quantities for BCDs from Papaderos 
et al., and additional dIs from Patterson \& Thuan.  {\it The host galaxies 
of BCDs lie at the extremes of the dwarf
irregulars in both plots}, in the sense that the BCDs have systematically
higher central surface brightnesses and smaller disk scale lengths.  That
is, the hosts of BCDs are significantly more compact, and possess higher
central mass densities than do more quiescent dIs.  We stress that our
measurements for the BCDs are done in a way that ignores the light from 
the starburst -- the quantities plotted are for the underlying host 
galaxies only.  If the light from the BCD were included, the difference
between the BCDs and dIs would be even greater.

\section{Discussion: Implications for Dwarf Galaxy Evolution}

The results of our surface photometry, when combined with the recent
studies mentioned in Section 3, can now be used to address the question 
posed in the introduction: are BCDs just typical low-surface-brightness 
dI galaxies currently in a bursting state?  We believe that the answer 
to that question is a definite NO!

The typical BCD appears to be hosted by a gas-rich galaxy with
structrual parameters at the extreme end of those exhibited by dwarfs.
On average, BCDs are hosted by centrally concentrated systems, with
very small scale lengths and high central mass densities.  If there
is a continuum of values for the scale length and central surface
brightness for dwarf galaxies, BCD hosts occupy the extreme end of the
distribution.  If this picture is correct, then BCDs cannot be hosted by
just any dwarf galaxy.  Rather, they are preferentially found in the
most centrally concentrated systems.  Further, they appear to occur
in systems with the highest central HI gas densities.  This
makes perfect sense, since it would be exactly these systems which
favor repeated star-formation bursts, and which could build up enough
gas mass in a restricted area to generate a star-formation episode of
sufficient magnitude to be called a BCD.

This scenario has a number of important implications for our views
on how dwarf galaxies evolve.  If BCDs and other dwarfs do not lose 
most of their gas mass as a result of a starburst, as suggested by the 
simulations of MacLow \& Ferrara (1998), they will not evolve into 
gas-poor dE galaxies, as some authors have contended.  Further, since 
little mass is ejected, there will be no significant dynamical
relaxation of the system.  In other words, the structural parameters
will not change significantly after the starburst is over.  Rather,
the post-burst BCDs will remain compact.  By remaining in such a
configuration, with high central mass densities and high HI gas content,
the period of time between bursts for a BCD could be quite short, 
probably on the order of the time scale for the cold ambient gas to
settle down after the effects of the last burst are over (perhaps on
a time scale of $\sim$100 Myr).  If correct, this implies a very high
duty cycle for BCDs.  It also helps to explain why we don't see
large numbers of compact, post-burst dwarf galaxies.  First of all,
most dwarfs cannot become BCDs in the first place, and those that
do might spend a large fraction of their time in a bursting phase.

However one decides to interpret the current results, it seems
clear that the old picture of dwarf galaxy evolution, where most or all
gas-rich dwarfs go through one or more BCD-like episodes, is not correct.
Rather, only a special subsample of the dwarf galaxy population can
host a starburst of the magnitude that would qualify it as a BCD.  This
simple result requires substantial rethinking of our view of how dwarfs
galaxies evolve, since in this scenario most galaxies do not evolve
through the BCD phase.  Thus transitions from gas-rich dIs to gas-poor
dEs seem much less likely, and the galaxies which are currently seen
as BCDs will not evolve into either normal dIs or gas-poor dEs after
the starburst phase is over.  A corollary to this is that galaxies
currently seen as low-surface-brightness dwarfs have most likely never 
gone through a BCD-like phase, and probably never will.

\acknowledgments

We are grateful to our many collaborators on our various dwarf galaxy
projects for input and suggestions.  In particular, thanks to Liese
van Zee, Katherine Rhode, David Westpfahl, and David Sudarsky.  JJS 
gratefully acknowledges financial support from the National Science
Foundation.

%


\begin{references}
\reference Brighenti, F., \& D'Ercole, A. 1999, in preparation
\reference De Young, D.S., \& Heckman, T.M.  1994, \apj, 431, 598
\reference MacLow, M.-M., \& Ferrara, A. 1998, \apj, in press
\reference Norton, S.A., \& Salzer, J.J.  1999, in preparation
\reference Papaderos, P., Loose, H.-H., Fricke, K.J., \& Thuan, T.X.
1996a, \aap, 314, 59
\reference Papaderos, P., Loose, H.-H., Thuan, T.X., \& Fricke, K.J.
1996a, \aaps, 120, 207
\reference Patterson, R.J., \& Thuan, T.X.  1996, \apjs, 107, 103
\reference Rhode, K.L., Salzer, J.J., \& Westpfahl, D.J.  1999, this volume
\reference Salzer,~J.~J., \& Elston,~R. 1992, in {\it I.A.U. Symposium 
No. 149}, ed. B. Barbuy and A. Renzini (Dordrecht: Kluwer), p. 482
\reference Salzer, J.J., Rosenberg, J.L., Weisstein, E.W., Mazzarella, J.M.,
\& Bothun, G.D.  1999a, in preparation
\reference Salzer,~J.~J., Sudarsky, D.L., \& Elston, R. 1999b, in preparation
\reference Searle, L. \& Sargent, W.L.W.  1972, \apj, 173, 25
\reference Telles, E., \& Terlevich, R. 1997, \mnras, 286, 183
\reference Thuan, T.X., Izotov, Y.I., \& Lipovetsky, V.A.  1997, \apj, 477, 661
\reference Thuan, T.X., \& Martin, G.E.  1983, \apj, 247, 823
\reference van Zee, L., Skillman, E.D., \& Salzer, J.J.  1998, \aj, 116, 1186
\end{references}
\end{document}